\documentclass[times, twocolumn]{aastex631}

\usepackage{amsmath, amssymb, bm}
\usepackage[version=4]{mhchem}
\usepackage{bbm}
\usepackage{verbatim}
\usepackage{soul}

\newcommand\tmod[1]{{#1}}

\shorttitle{Atomic Jet of TMC1A}
\shortauthors{Harsono et al.}
\begin{document}

\title{JWST Peers into the Class I Protostar TMC1A: Atomic Jet and Spatially Resolved Dissociative Shock Region}

\email{dharsono@gapp.nthu.edu.tw}

\author[0000-0001-6307-4195]{D. Harsono}
\affiliation{Institute of Astronomy, Department of Physics, National Tsing Hua University, Hsinchu, Taiwan}

\author[0000-0002-7993-4118]{P. Bjerkeli}
\affiliation{Chalmers University of Technology, Department of Space, Earth and Environment, SE-412 96 Gothenburg, Sweden}

\author[0000-0002-3835-3990]{J. P. Ramsey}\affiliation{Department of Astronomy, University of Virginia, Charlottesville, VA 22904, USA}

\author{K. M. Pontoppidan}
\affiliation{Space Telescope Science Institute, 3700 San Martin Drive, Baltimore, MD 21218, USA}

\author[0000-0003-1159-3721]{L. E. Kristensen}
\affiliation{Niels Bohr Institute, University of Copenhagen, {\O}ster Voldgade 5--7, DK~1350 Copenhagen K., Denmark}

\author{J. K. J{\o}rgensen}
\affiliation{Niels Bohr Institute, University of Copenhagen, {\O}ster Voldgade 5--7, DK~1350 Copenhagen K., Denmark}

\author[0000-0003-3393-294X]{H. Calcutt}
\affiliation{Institute of Astronomy, Faculty of Physics, Astronomy and Informatics, Nicolaus Copernicus University in Toru{\'n}, ul. Grudzik{a}dzka 5, 87--100 Toru{\'n}, Poland}

\author[0000-0002-7402-6487]{Z-Y. Li}
\affiliation{Department of Astronomy, University of Virginia, Charlottesville, VA 22904, USA}

\author[0000-0002-9912-5705]{A. Plunkett}
\affiliation{National Radio Astronomy Observatory, 520 Edgemont Road, Charlottesville, VA 22903, USA}

% DOI: 10.17909/6k3w-1948

\begin{abstract}

Outflows and winds launched from young stars play a crucial role in the evolution of protostars and the early stages of planet formation. However, the specific details of the mechanism behind these phenomena, including how they affect the protoplanetary disk structure, are still debated. We present {\it JWST} NIRSpec Integral Field Unit (IFU) observations of atomic and \ce{H2} lines from 1 -- 5.1 $\mu$m toward the low-mass protostar TMC1A. For the first time, a collimated atomic jet is detected from TMC1A in the [\ion{Fe}{2}] line at 1.644 $\mu$m along with corresponding extended \ce{H2} 2.12 $\mu$m emission. Towards the protostar, we detected spectrally broad \ion{H}{1} and \ion{He}{1} emissions with velocities up to 300 km/s that can be explained by a combination of protostellar accretion and a wide-angle wind. The 2$\mu$m continuum dust emission, \ion{H}{1}, \ion{He}{1}, and \ion{O}{1} all show emission from the illuminated outflow cavity wall \tmod{and scattered line emission}. These observations demonstrate the potential of {\it JWST} to characterize and reveal new information about the hot inner regions of nearby protostars. In this case, a previously undetected atomic wind and ionized jet in a well-known outflow.

% Abstract: 250 words, text: 3500 words, no more than 5 figures and tables

\end{abstract}


\section{Introduction}\label{sec:intro}

The formation and growth of a protostellar system require material to collapse and accrete onto a central mass via gravity, and these effects are observable as infalling motion \citep[e.g.][]{SL77,leung77} and accretion \citep[e.g.][]{herbig77,hartmann85}. However, the most visually prominent manifestation of ongoing star formation is bipolar outflows that can be traced through atomic \citep[e.g.][]{mundt83, persson84} and molecular \citep[e.g.][]{bally83, snell90, davis02} gas lines.  Outflows are an important part of any protostellar system since they remove angular momentum from the protoplanetary disk that orbits the forming protostar. This \tmod{further} aids the accretion process from the disk toward the protostar \citep[e.g.][]{frank14}. The removal of material by the outflow may also affect the physical structure of the disk, and therefore potentially influence the disk evolution, with consequences for any planets that eventually form in the disk \citep[e.g.][]{pascucci22}.

Protostellar outflows are often described as consisting of a collimated jet and a slower wide-angle wind, even though distinctions between the components and their physical origins are debated \citep{arce07}.  The jet component denotes the fast and collimated outflow that is likely responsible for entrainment on scales $>5000$ au as observed through, e.g., CO emission \citep{rabinanahary22, schutzer22}. Meanwhile, a slower wide-angle disk wind points to the small-scale ($< 500$ au) outflow which is closely related to the protoplanetary disk. Observationally, both jet and wide-angle wind components should be detectable in the IR and sub-mm. Protostellar jets are observed through high velocity $\gtrsim 100$ km/s rotational transitions of \ce{^{12}CO} and SiO in the sub-mm \citep[e.g.,][]{tafalla04,hirano10, plunkett15, hull16, ray21} in addition to shocked gas tracers in the infrared such as [\ion{O}{1}] 63 $\mu$m, [\ion{Fe}{2}] 1.64 $\mu$m, and \ce{H2} 2.12 $\mu$m lines \citep[e.g.,][]{hartigan94, nisini04,gusdorf17, yang22}. The slower disk wind is typically seen using rotational transitions of \ce{^{12}CO}, and SO in the sub-mm \citep{podio15, bjerkeli2016, tabone17, devalon20} along with \ion{O}{1} lines in the IR \citep{pascucci22}.

The Class I protostar TMC1A (IRAS 04365+2535) is an embedded $\sim$0.45 $M_\odot$ \citep{harsono18a} protostar in the Taurus Molecular cloud \citep[140 pc,][]{galli19}. It has a large-scale bipolar \ce{^{12}CO} molecular outflow that extends to at least 6000 au \citep{yildiz15}. Atacama Large Millimeter/submillimeter Array (ALMA) observations of the $J=2-1$ rotational transition of \ce{^{12}CO} at 1.3mm \citep{bjerkeli2016} show a molecular disk wind emanating from the protoplanetary disk. Fast-moving gas is also seen in ground-based, high-spectral resolution ($\sim 3$ km/s) observations of ro-vibrational lines of \ce{^{12}CO} (4.6 $\mu$m) that show hot ($>1000$ K) molecular gas moving at radial velocities of $\sim 75$ km/s ($\sim$ 93 km/s when accounting for an inclination of $i = 54^{\circ}$, \citealt{harsono21}), indicative of a disk wind \citep{herczeg11}. Despite the detections of the slow disk wind component, a jet component of TMC1A has not previously been detected in SiO \citep{harsono21} or water lines \citep{kristensen12}.

 {\it JWST} provides the spectral ($R = 2700$) and spatial resolution 
(0\farcs{1}) needed to better understand the nature of the outflow around TMC1A. 
 With the Near Infrared Spectrograph (NIRSpec; \citealt{jwstnirspec}) 
instrument, the hot ($T_{\rm gas} > 1000$ K) atomic and molecular gas has been 
mapped in the vicinity of TMC1A at wavelengths between 0.9--5 $\mu$m.  In 
this letter, we present the first detection of a jet through [\ion{Fe}{2}] 
emission.  We also report and analyze the hot atomic and molecular lines from 
\ion{H}{1}, \ion{He}{1}, \ion{O}{1}, [\ion{Fe}{2}], and \ce{H2} coming from the 
inner regions of the protostar. In combination with archival data from other 
observatories, these results provide the means to start building a comprehensive 
picture of the TMC1A protostellar + outflow system.

%%%%%%%%%%%%%%%%%%%%%%%%%%%%%%%%%%%%%%%%%%%%%%%%%%%%%%%%%%%%%%%%%%%%%%%%%%%
%%

\section{JWST observations and calibration} \label{sec:jwstobs}

{\it JWST} observed TMC1A on September 2 2022 using the NIRSpec integral field unit (IFU) mode (PID: 2104, PI: Harsono).  {\it JWST} NIRSpec provides high-spectral resolution data from $\sim 0.97$ $\mu$m to 5.28 $\mu$m, nominally across a 30$\times$30 grid of 0\farcs{1} ``spaxels" (14 au at 140 pc).  The F100LP (1 -- 1.9$\mu$m), F170LP (1.7 -- 3.2 $\mu$m), and F290LP  (2.9 -- 5.3 $\mu$m) filters were used with G140H, G235H, and G395H dispersers for the highest spectral resolution ($R \sim 2700$ or $\Delta v = 90 - 150$ km/s; \citealt{jwstnirspec}).  The exposure times were 15719 s, 4123 s, and 2577 s for F100LP, F170LP, and F290LP filters, respectively.  Target acquisition was performed with Wide Aperture Target Acquisition (WATA) using the full sub-array in the F110W filter, and data was obtained using the NRSRAPID readout pattern with an exposure time of 42.9 seconds.  The science observations were taken with a medium CYCLING dither pattern and a total of 6 dithers for each filter.  The science integration was taken with 4, 15, and 60 groups in the F100LP, F170LP, and F290LP filters.  For this letter, we concentrate on the gas lines in the wavelength range between 0.97 $\mu$m to $\sim2$ $\mu$m where we detect many atomic lines, and we will present the complete data in a follow-up work.

\tmod{
We re-processed the Stage 2 products as processed through the {\it JWST} pipeline, version 1.9.6 \citep{jwstpipe} with the calibration files CRDS \textsc{jwst\_1088.pmap} \citep{jwstcrds}. For the production of IFU spectral cubes, additional rejections were applied to remove the outliers in stage 2 products before passing them through the stage 3 step of the pipeline. 
} 
A flux-calibrated spectrum has been extracted from the central square aperture of 0\farcs{3} centered on the brightest spaxel. The systematic variation in the velocity resolution is approximately 50--100 km/s (1--2 channels).

We used the National Institute of Standards and Technology (NIST) atomic spectra database \citep{kramida10} and the atomic line list of \citet{vanhoof18} to obtain the properties of transitions for H, He, O, and Fe. A local average of the dust continuum emission is subtracted from each emission line before creating zeroth moment maps. The statistical noise of each map is calculated from the emission-free regions (spatially and spectrally). In addition to the JWST observations, we include archival ALMA observations \citep[][PID:2017.1.00212.S]{bjerkeli2016} and {\it Hubble Space Telescope} (HST) observations with the Near Infrared Camera and Multi-Object Spectrometer (NICMOS) in the F160W filter (PID:7325, PI: Terebey). These data are used to align the images (see App.~\ref{app:align}) and to compare the emission morphology at different wavelengths.

\begin{figure}
    \centering
    \includegraphics[width=0.45\textwidth]{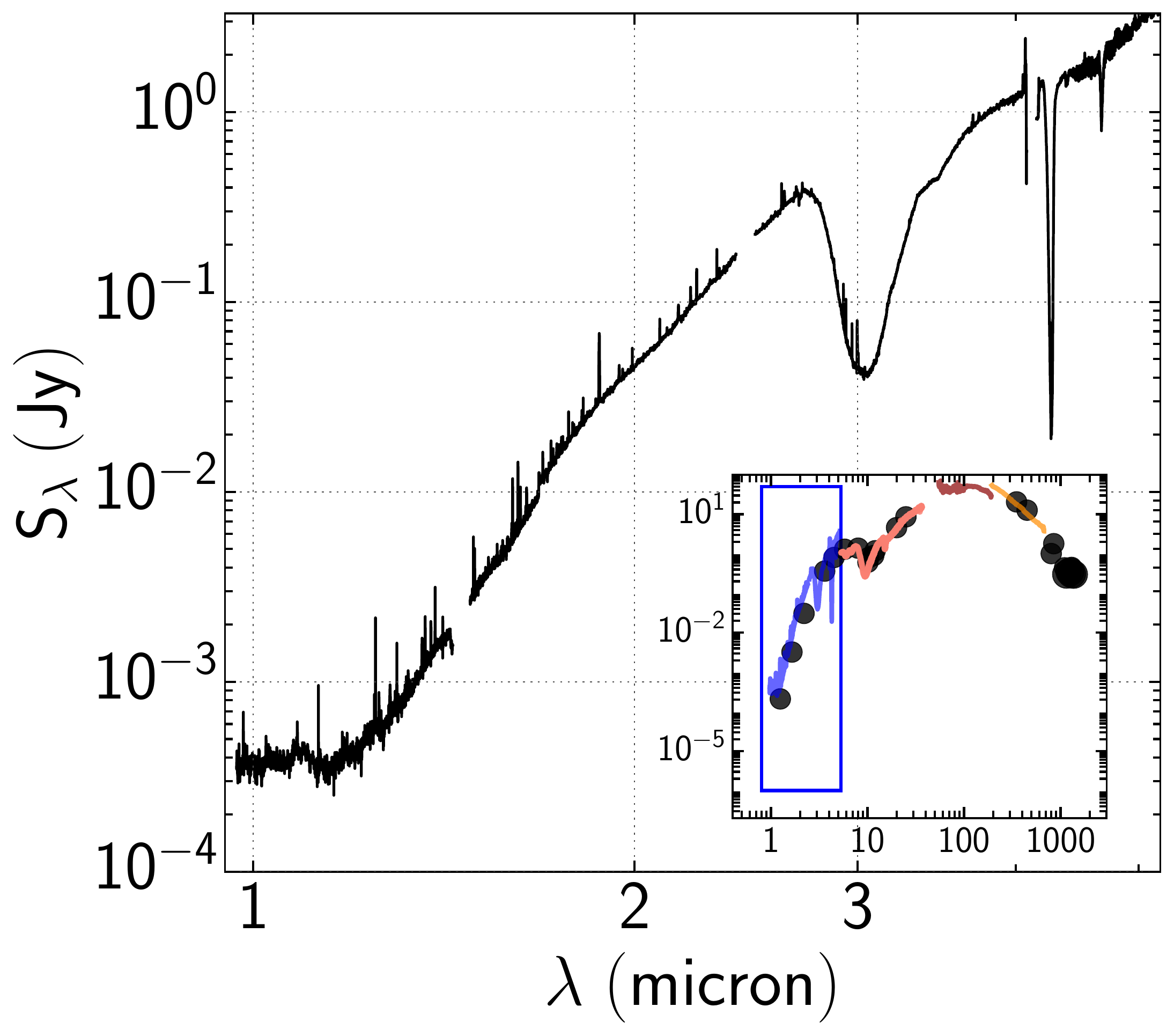}
    \caption{NIRSpec spectrum extracted from the central 0\farcs{3} spaxels. The spectral energy distribution of TMC1A from other observations is shown in the inset along with the NIRSpec data in blue.  The Photometric data points (black) are from \citet{kristensen12} (includes points from \citealt{skrutskie06}, \citealt{francesco08},  \citealt{evans09}, \citealt{karska13}, and \citealt{jdgreen13}) .  {\it Herschel} PACS observations of \citet{jdgreen13} are shown in orange. The {\it SPITZER} IRS spectrum \citep[pink;][]{lahuis10} is overlaid.}
    \label{fig:tmc1ased}
\end{figure}

%%%%%%%%%%%%%%%%%%%%%%%%%%%%%%%%%%%%%%%%%%%%%%%%%%%%%%%%%%%%%%%%%%%%%%%%%%%
%%

\section{Results} \label{sec:results}

\begin{figure*}
    \centering
    \includegraphics[width=0.9\textwidth]{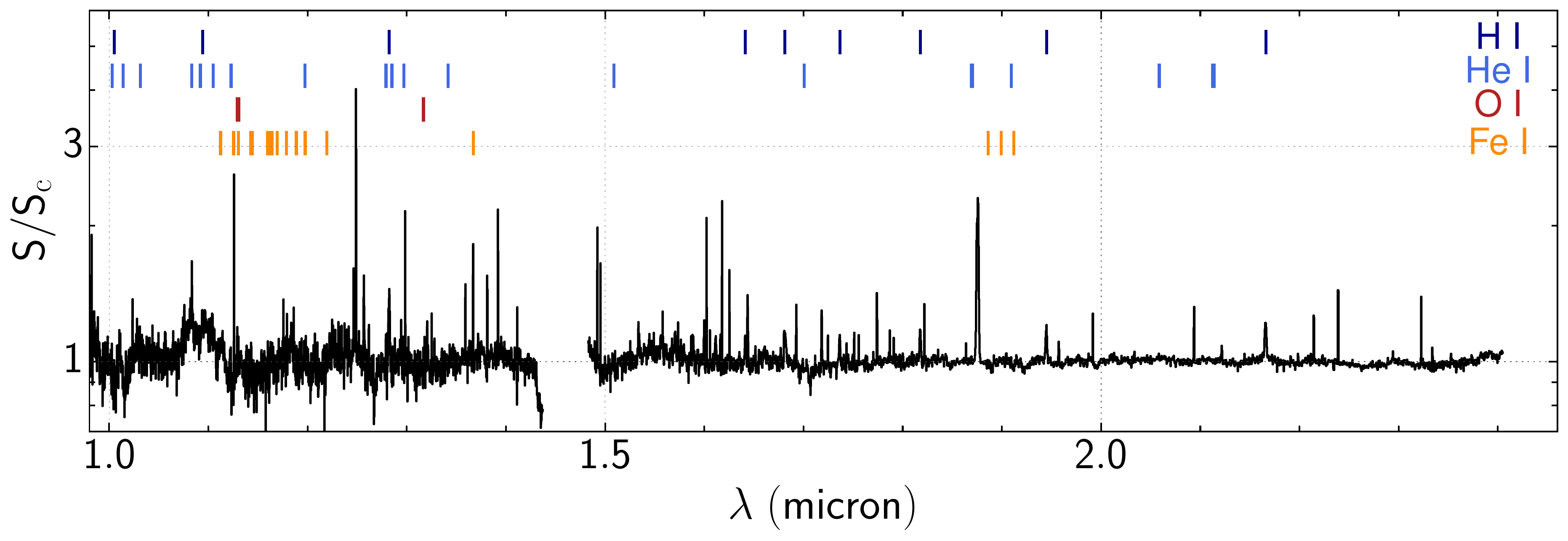}
    \includegraphics[width=0.95\textwidth]{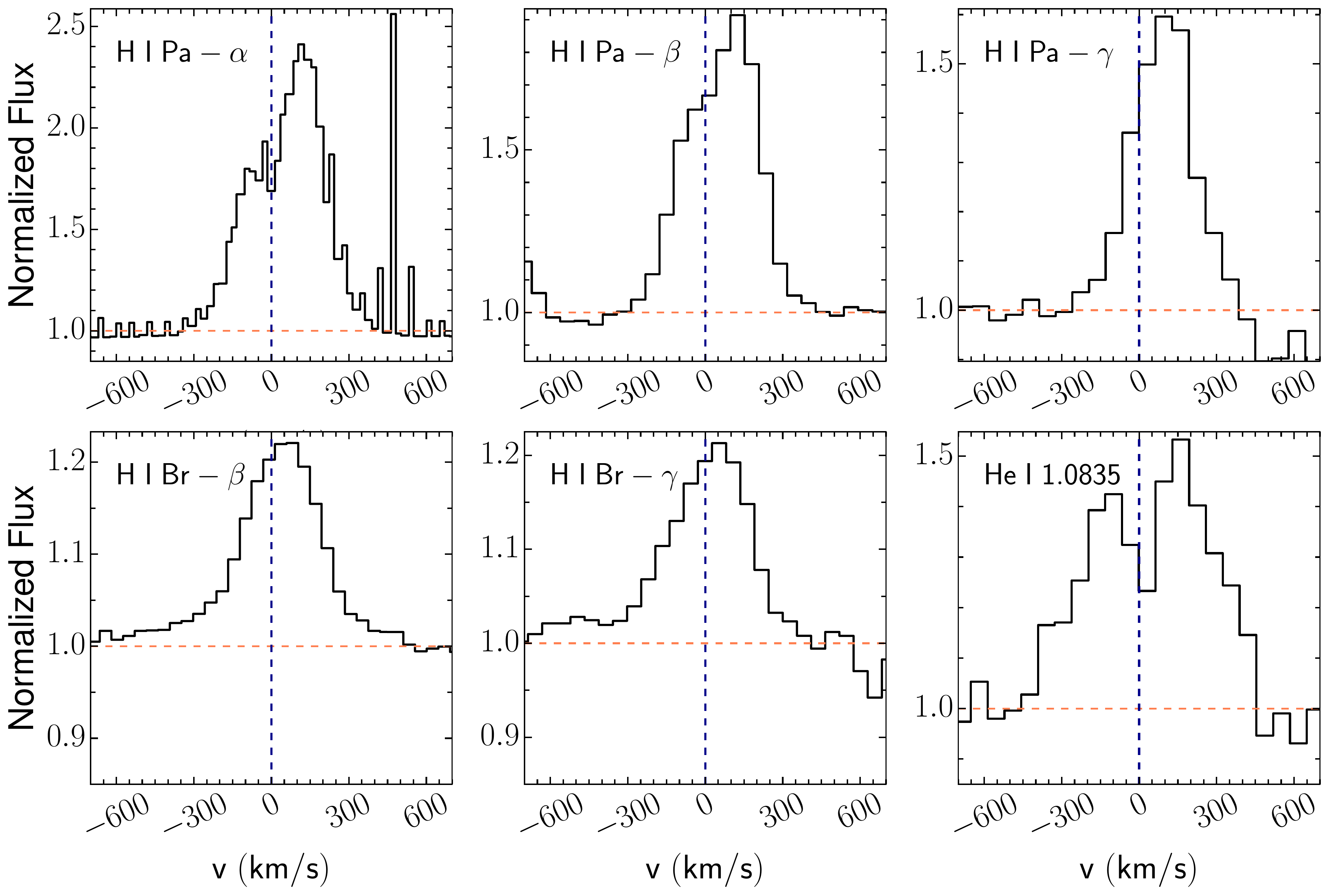}
    \caption{{\it Top:} Normalized spectrum of TMC1A. The spectrum has been normalized with a global continuum fit constructed with a polynomial of degrees 5 to 7.  Transitions of \ion{H}{1}, \ion{He}{1}, \ion{O}{1}, and \ion{Fe}{2} with high $A_{\rm ij}$ values are indicated by the vertical colored lines above the spectrum. {\it Bottom:} Zoom-in spectra of the brightest \ion{H}{1} lines and the \ion{He}{1} $\lambda$1.0835 line. The spectral resolution is $\Delta v \sim 100 - 150$ km/s.   }
    \label{fig:spectra}
\end{figure*}

The observed near-infrared spectrum from 1--5 $\mu$m towards TMC1A is shown in Fig.~\ref{fig:tmc1ased}.  The flux density spans $>$4 orders of magnitude between a few micro Jansky at $\sim 1$ $\mu$m and $\sim$ 0.2 Jy at $\sim 2.8$ $\mu$m.  The complete spectral energy distribution (SED) is also shown in the inset of Fig.~\ref{fig:tmc1ased} to indicate where our spectrally resolved observations fit in. The photometric data are retrieved from \citet{kristensen12} and the spectrum from the {\it SPITZER} "Cores to Disks" (c2d) program  \citep{lahuis10}.  Flux densities in the $70$ $\mu$m to 200 $\mu$m range are taken from the {\it Herschel} "Dust, Ice, and Gas in Time" (DIGIT) program \citep{jdgreen13}.  By including {\it JWST}, TMC1A now has continuous spectral coverage at moderately high spectral resolution from $\sim 1$ $\mu$m up to $\sim$200 $\mu$m. In this paper, we only analyze the spectrally resolved data toward the brightest spaxel.

Figure~\ref{fig:spectra} shows the normalized spectrum with lines from \ion{H}{1}, \ion{He}{1}, \ion{O}{1}, and \ion{Fe}{2} as indicated. The spectrum was normalized using piece-wise polynomials of degrees 5 to 7 fitted to the line-free channels.  Numerous gas lines are detected down to $\sim 1$\% of the continuum but we concentrate on the brightest atomic and molecular lines only for this study.

Figure~\ref{fig:spectra} (bottom) shows the spectrally resolved \ion{H}{1} Paschen series ($n_{\rm final}$ = 3) and Brackett series ($n_{\rm final}$ = 4) lines which are detected at high $S/N$ ($>10$ based on the statistical noise around each line).  The \ion{H}{1} lines in the Paschen series show an asymmetric, redshifted line profile with peaks of the different lines located at $v\sim 100$ km/s.  The broad, red asymmetric line profile could be indicative of wind emission, similar to previously detected ro-vibrational fundamental mode ($v = 1-0$) lines of CO \citep[see App.~\ref{app:rovib},][]{herczeg11}. The Brackett lines are similarly broad, but less asymmetric, profiles with the peak intensity of the lines shifting to a slower velocity of $\sim$50 km/s.

\begin{figure*}
    \centering
    \includegraphics[width=0.97\textwidth]{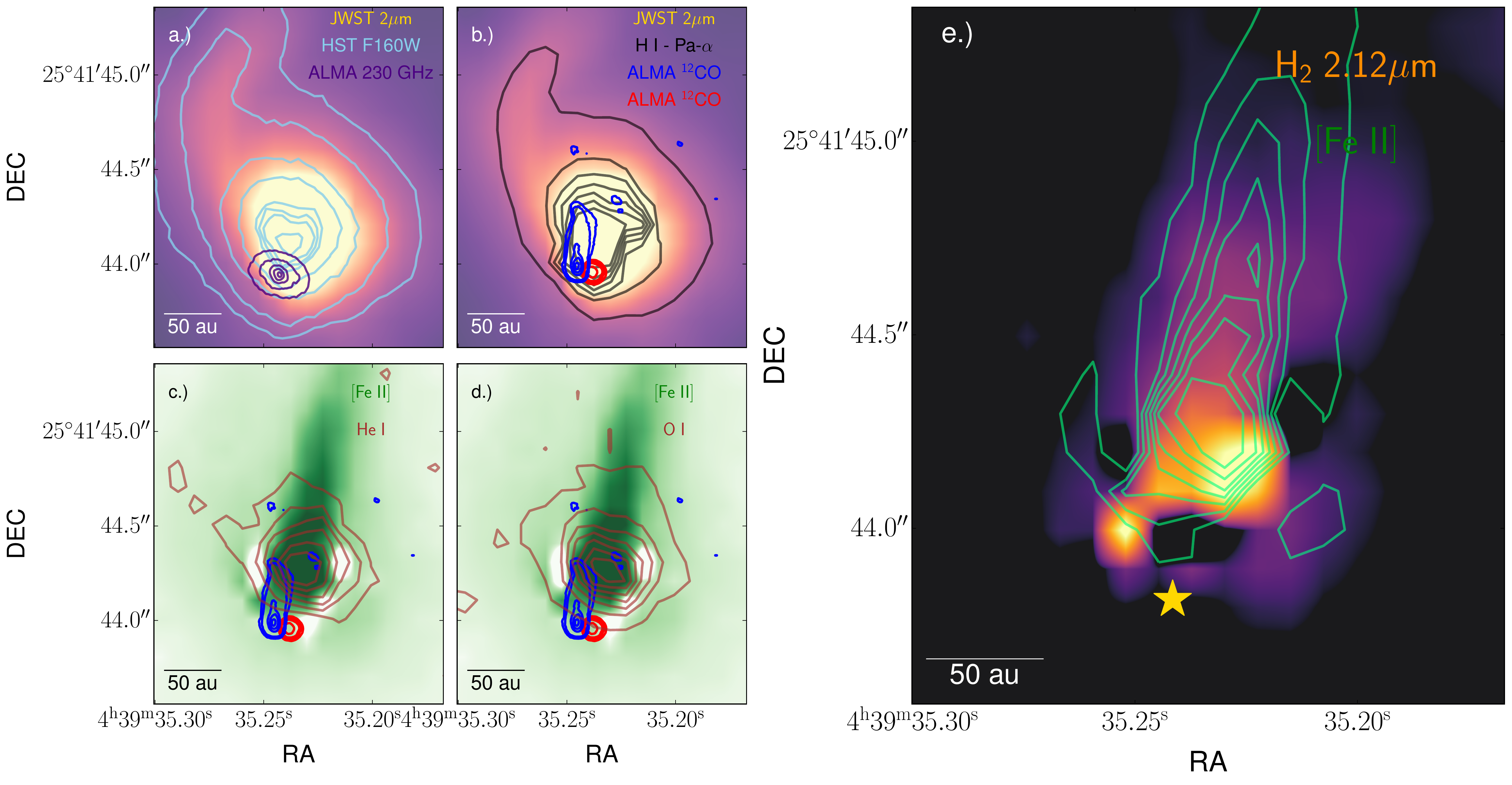}
    \caption{{\it a.)} {\it JWST} 2$\mu$m continuum map is shown in color. A linear scale between 0.5\% to 99\% of the intensity distribution is used. We have overlaid the {\it HST} F160W and ALMA 230 GHz continuum maps on top of the 2$\mu$m map to orient our results with respect to previous observations. The contours are chosen to highlight the features of the {\it HST} image and the location of the millimeter dust disk. 
    {\it b.)} The integrated \ion{H}{1} Pa-$\alpha$ map is shown by the black contour lines over top of the 2$\mu$m map. Contours from 95\% to 98\% of the intensity distribution are used. ALMA observations of the blue- and red-shifted rotational transition of \ce{^{12}CO} ($J=2-1$) are shown by the blue and red contours, respectively. The \ce{^{12}CO} $J=2-1$ line is only showing velocities $>$ 1.5 km/s from systemic. The contours are chosen to highlight the orientation of the \ce{^{12}CO} $J=2-1$ gas. 
    {\it c.)} A comparison between [\ion{Fe}{2}] 1.644 $\mu$m, \ion{He}{1} $\lambda$10835, and \ce{^{12}CO} $J=2-1$ line emission. The integrated [\ion{Fe}{2}] line is shown in the background in green with a linear scale spanning between 60\% to 99\% of the intensity distribution. The \ion{He}{1} map is shown by the brown contours with levels from 95\% to 97\% of the intensity distribution). 
    {\it d.)} A comparison between [\ion{Fe}{2}] 1.644 $\mu$m, \ion{O}{1} $\lambda$1.12, and \ce{^{12}CO} $J=2-1$ lines. The contours of \ion{O}{1} (in brown) are also showing the brightest components (95\%--97\% of the intensity distribution). 
    {\it e.)} The integrated [\ion{Fe}{2}] map is overlaid on top of the integrated \ce{H2} 2.12$\mu$m map. The \ce{H2} map is drawn with colors spanning from 5$\sigma$ up to the peak intensity. The green contours show [\ion{Fe}{2}] but only the brightest pixels in the map ($>96$\% of the intensity distribution). The center of the millimeter dust disk is indicated by the star. 
    In each panel, a 50 au scale bar is show in the bottom left. 
    }
    \label{fig:bigcombine}
\end{figure*}

\tmod{\ion{H}{1} lines are thought to primarily trace the magnetospheric accretion flows and/or partially ionized disk winds} \citep{sedwards06, erkal22}. By integrating the \ion{H}{1} lines between -400 to 400 km/s, we find equivalent widths ($EW$) of the Pa-$\alpha$ and Pa-$\beta$ lines between 300 to 550 km/s ($\sim10$--35 \AA).  The measured line strengths are similar to previous \ion{H}{1} studies toward the outflows from classical T-Tauri disks \citep{muzerolle98, sedwards13}.  Paschen and Brackett line ratios can be used to estimate the density and temperature of the gas.  We find that the integrated line ratios for TMC1A are Pa-$\alpha$/Pa-$\beta \simeq 1.8$, Pa-$\beta$/Pa-$\gamma \simeq 1.6$, and Br-$\gamma$/Pa-$\beta \simeq 0.27$.  Using the models of \citet{kwan11}, these line ratios indicate emission from hot dense gas with $n_{\rm H} > 10^{9}$ cm$^{-3}$ and $T_{\rm gas} > 5000$ K.

\tmod{Assuming that the line emission is dominated by magnetospheric accretion, the \ion{H}{1} line luminosity can be used to infer the stellar accretion}.  Since the measured $EW$ is similar to the classical T-Tauri disks, we follow \citet{muzerolle98} to estimate the accretion rate of TMC1A.  The measured integrated $EW_{\rm Pa-\beta}$ = 305 km/s translates to $L_{\rm Pa-\beta} = 7.17 \times 10^{-8} \ L_{\odot}$.  Using the $L_{\rm Pa-\beta} - L_{\rm acc}$ relation, we obtain $\dot{M} \sim 2 \times 10^{-12} \ M_{\odot}$ yr$^{-1}$, which is much lower than expected for a Class I source.  Meanwhile, the Br-$\gamma$ line (2.1661 $\mu$m) has an $EW \sim 73$ km/s, which translates to an accretion rate of $\dot{M} \sim 3 \times 10^{-9}$ $M_{\odot}$ yr$^{-1}$. Note that both line ratios and accretion rates are calculated without accounting for any extinction.

Based on the bolometric luminosity (2.7 $L_{\odot}$, \citealt{kristensen12}) and the protostellar mass (0.45 $M_{\odot}$, \citealt{harsono21}), the expected accretion rate is $\sim 1 \times 10^{-6}$ $M_{\odot}$ yr$^{-1}$. \tmod{ A rate of $\sim 5.6 \times 10^{-6}$ $M_{\odot}$ yr$^{-1}$ can be estimated using the envelope density model of \citet{furlan08}.}  Through the comparison of these two accretion rates, the upper limit to the visual extinction is $A_{\rm v} = 10-30$ ($A_{\rm J} = 0.3 \ A_{\rm V}$, \citealt{pontoppidan04}).  Meanwhile, the measured \ion{H}{1} Pa-$\beta$/Pa-$\gamma$ and  Br-$\gamma$/Pa-$\beta$ ratios suggests $A_{\rm V} < 5$ as compared to \citet{sedwards13}.  With the spatial information afforded by the NIRSpec IFU, it is possible to constrain how the extinction towards TMC1A varies with position, but this is left to a future study.

We also detect and spectrally resolved the \ion{He}{1} 1S-1Po $\lambda10835$ line (Fig.~\ref{fig:spectra}). The line is double-peaked with a similar line width as the \ion{H}{1} Pa-$\alpha$ line. This \ion{He}{1} line can also be used to trace magnetospheric accretion and the disk wind \citep{sedwards06, erkal22}. The deficit of blue-shifted absorption may indicate that the line emission originates from the dense jet and disk wind with a minor contribution from magnetospheric accretion.  Meanwhile, the \ion{O}{1} and [\ion{Fe}{2}] lines are narrow with typical line widths of 1--2 channels.

The zeroth moment maps of the atomic and \ce{H2} 2.12 $\mu$m lines can be used to highlight the physical components of TMC1A on 50 au scales which are shown in Fig.~\ref{fig:bigcombine}. In order to orient the data with respect to the disk and \ce{^{12}CO} wind, we overlay the {\it HST} and ALMA 230 GHz dust continuum data on top of the {\it JWST} 2$\mu$m continuum map in panel (a). The second panel (b) shows the extended \ion{H}{1} Pa-$\alpha$ line with a similar structure as the 2$\mu$m dust continuum. Note that the cold molecular wind ($T_{\rm gas} \sim 60$ K) that is traced by the \ce{^{12}CO} $J=2-1$ line with ALMA is inside of the \ion{H}{1} emission. Panels (c) and (d) show the \ion{He}{1} and \ion{O}{1} zeroth moment maps overlaid on top of the [\ion{Fe}{2}] emission map. Blue and red contours continue to mark the sub-mm \ce{^{12}CO} wind. The [\ion{Fe}{2}] emission is extended while both He and O line emission concentrates around the dust continuum peak. Finally, the [\ion{Fe}{2}] and the v=1--0 S(1)\ce{H2} integrated maps are shown in panel (e) where they show a similar extended structure. The peak of the [\ion{Fe}{2}] emission is slightly offset from \ce{H2}.

%%%%%%%%%%%%%%%%%%%%%%%%%%%%%%%%%%%%%%%%%%%%%%%%%%%%%%%%%%%%%%%%%%%%%%%%%%%
%%

\begin{figure}
    \centering
    \includegraphics[width=0.5\textwidth]{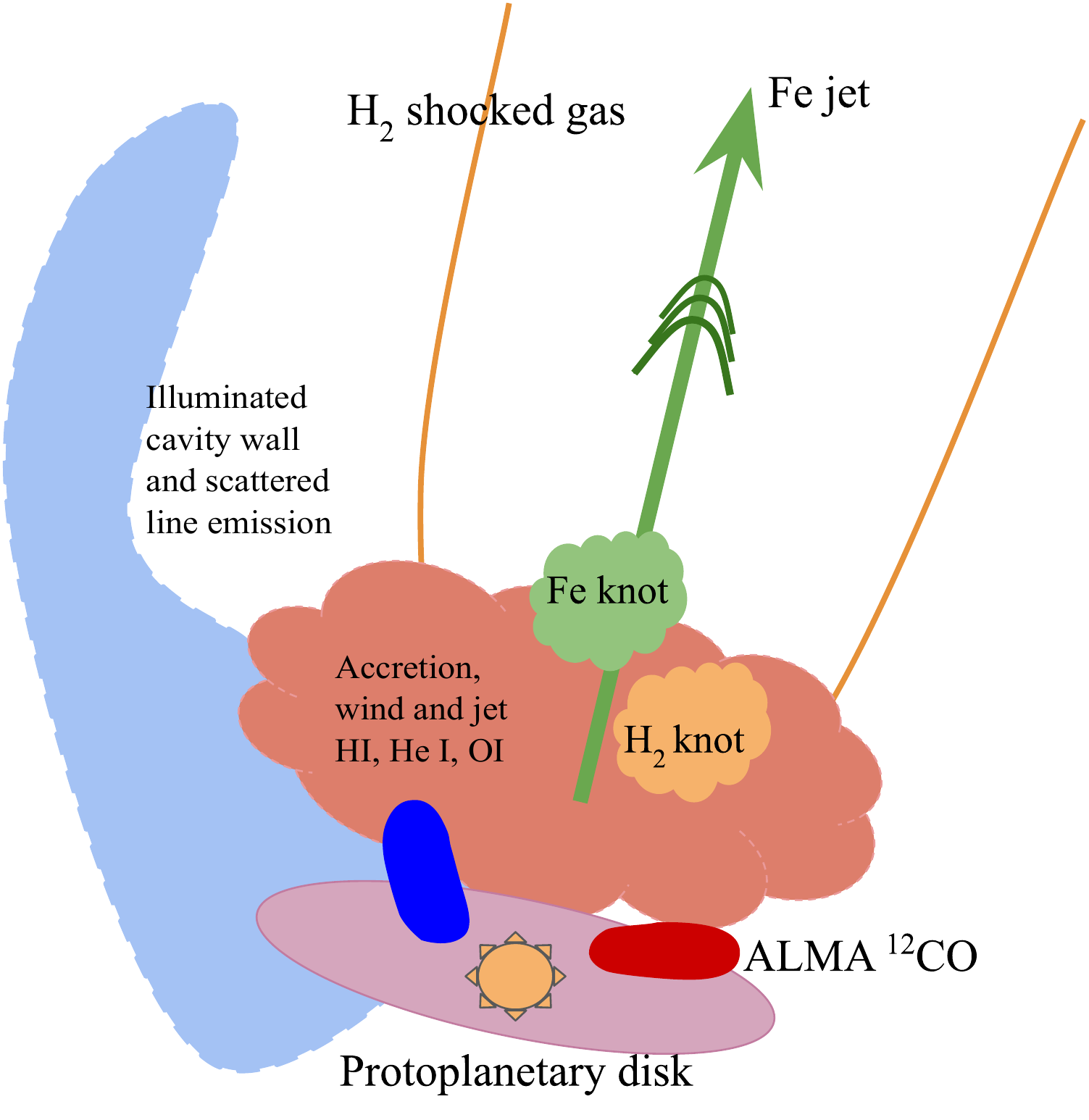} 
    \caption{Cartoon of TMC1A's physical structure. The three major components that are traced by the hot gas lines are accretion to the protostar, the jet, and the illuminated cavity wall. The disk and the cold wind as traced with ALMA are also indicated. 
    }
    \label{fig:cartoon}
\end{figure}

\section{Discussion} \label{sec:disc}

Our {\it JWST} NIRSPEC observations show spectrally resolved atomic and molecular lines between 1--5 $\mu$m. We detected broad atomic emission lines ($>200$ km/s) from the central 0\farcs{3} region. With the IFU, we also see linearly extended emission in the direction of the outflow traced by [\ion{Fe}{2}] and \ce{H2}. Finally, we detect diffuse atomic emission on-source that overlaps with the NIR continuum emission. We will discuss the origin of each observable feature below and the implications \tmod{for} protostellar accretion and outflows.

In contrast to the more evolved Class II disks, the broad \ion{H}{1} and \ion{He}{1} line profiles observed toward TMC1A indicate a significant contribution from the jet and wind in addition to the magnetospheric accretion \citep{nisini04,sedwards06}. Magnetospheric accretion models tend to produce narrow ($\Delta v \sim$ 100 km/s) line profiles with an inverse P-Cygni absorption \citep{muzerolle98b, kurosawa06}. Given how broad our line profiles are, it is likely that the \tmod{outflow (jet, entrained gas, and the wind)} is the dominant source of emission at the brightest spaxel as presented in this letter.  In accretion or disk wind scenarios, the observed emission comes from the largely neutral gas that is either collisionally excited or recombination lines.  The observed Pa-$\beta$/Pa-$\gamma$ line ratio is similar to that of the luminous Class I protostar HH100 IRS \citep[$>1.4$;][]{nisini04}, which is consistent with a partially ionized gas. In order to estimate the physical conditions of the gas from the observed line ratios, we use the \citet{kwan11} model that includes the local excitation of \ion{H}{1}. The line ratio indicates optically thick ($\tau \sim 10$) \ion{H}{1} emission with a column density between $10^{10}$ and $10^{12}$ cm$^{-2}$.  If the temperature of the gas is closer to 5000 K, a higher density of \ion{H}{1} is needed to explain the observed line ratio. Toward TMC1A, the ro-vibrational v=2--1 line of \ce{^{12}CO} also exhibits a broad profile ($\Delta v \sim 96$ km/s) similar to HH100 IRS with temperatures of $\sim 1000$ K \citep{herczeg11}. If both lines probe a similar region within the central 0\farcs{1} -- 0\farcs{2} (14--28 au), the data is tracing the high-density gas with temperatures $>1000$ K caused by both accretion and outflow.

The extended [\ion{Fe}{2}] and \ce{H2} emission correspond to the jet that drives the large-scale CO outflow.  Emission from [\ion{Fe}{2}] tends to trace dissociative shocks along the jet \citep{allen93} with velocities $\gtrsim 40$ km/s \citep{may20} at densities of $\sim 10^{4}$ cm$^{-3}$ \citep{mccoey04}.  [\ion{Fe}{2}] is expected to emit from internal shocks of the jet where faster material catches up and shocks against slower-moving material that has been ejected earlier.  Meanwhile, \ce{H2} emission is expected to trace the slower shock component behind the knots. Figure \ref{fig:bigcombine} shows that \ce{H2} emission peaks are offset by 30--40 au from [\ion{Fe}{2}].  It is usually suggested that the extended \ce{H2} shocked gas traces the wide-angle wind that surrounds the jet \citep{davis02, amboage14} that is responsible for opening the outflow cavity. Toward TMC1A, \ce{H2} emission is concentrated in a small conical region around the Fe jet while the \ion{H}{1} emission is more extended \tmod{as their emission is scattered by dust grains}. \ce{H2} emission seems to be clumpy near the upper layers of the disk that is caused by either extinction or the physical conditions surrounding the jet launching region.  In general, \ce{H2} emission can be explained by either \ce{H2} reformation in the wake of the leading bow shock of a jet or the presence of \ce{H2} in the shell around the [\ion{Fe}{2}] jet.

The origins of protostellar outflows are still debated, and two physically 
related launching mechanisms have emerged as the primary explanations: The 
wide-angle magneto-hydrodynamic (MHD) wind \citep[e.g.][]{pudritz83, mundt83} 
and the X-wind model \citep[e.g.][]{shu94, shang07}.  The wide-angle 
wind (disk-wind) is launched from the protoplanetary disk's surface while an 
X-wind is launched from a narrow region where the inner, dust-free disk's 
($T_{\rm dust} > 1400$ K) magnetic field and the protostar's magnetosphere 
interact. In future analyses, we will combine multiple atomic and \ce{H2} lines 
to assess whether or not the protostellar outflow from TMC1A is consistent with 
the proposed launching mechanisms.

The outflow cavity wall is traced by the dust continuum emission as seen in {\it HST} and {\it JWST}\ 2$\mu$m images. These structures are visible due to the scattering of light by micron-sized grains along the cavity wall.  The arc that is highlighted by the 2$\mu$m continuum and \ion{H}{1} line emission could be due to the local overdensities of micron-sized dust grains as a result of interactions with the wind and protostellar radiation, potentially even an illuminated accretion stream \tmod{, or scattered  line emission}. The opening angle of the outflow cavity as traced by the dust emission appears larger than the extent of the \ce{H2} emission. The presence of \ion{H}{1} along the arc could be \tmod{tracing} by the dissociation of \ce{H2} along the cavity wall by UV photons or shocks near the cavity produced by a disk- or X-wind. In the X-wind picture, the dissociation comes from the shocked gas as the wind interacts with the infalling envelope \citep[e.g.,][]{shang07,shang20}. With the slower disk wind, additional UV radiation from the protostar is needed to dissociate the molecular material along the cavity wall \citep{kristensen13}. The main criticism against fast, steady X-winds is that they would disrupt the protostellar envelope within $\sim$1000 years given the observed protostellar envelope structures around low-mass stars \citep{liang20}.  From the outflow of TMC1A and the disk-to-stellar mass ratio \citep{harsono21}, the protostellar age of TMC1A is $>10^{4}$ years. \tmod{Alternatively, the extended \ion{H}{1} emission is tracing the unresolved emission from the magnetospheric accretion and/or the wind as the line emission is forward-scattered by the dust grains along the arc. Therefore, the \ion{H}{1} line emission is as extended as the dust arc. } These observations are therefore more consistent with an extended disk wind model carving out the outflow cavity that is directly illuminated by the feedback (radiation and mechanical) from the accretion and outflow.

{\it JWST} is shedding light on the anatomy of the inner hot regions around TMC1A. Combined with  constraints from ALMA data, in Fig.~\ref{fig:cartoon}, we present a cartoon of the different physical components as traced by the different emission lines shown. The components are a planet-forming disk, accretion, jet/wind, and outflow cavity wall. With NIRspec, we now have access to high-spatial resolution observations of atomic and molecular tracers that can trace accretion and ejection directly along with their associated feedback on the immediate surroundings, which is \tmod{a} crucial information for improving our understanding of how stars form.

%%%%%%%%%%%%%%%%%%%%%%%%%%%%%%%%%%%%%%%%%%%%%%%%%%%%%%%%%%%%%%%%%%%%%%%%%%%%%%%%%%%%%%%%%%%
%%%%%%%%%%%%%%%%%%%%%%%%%%%%%%%%%%%%%%%%%%%%%%%%%%%%%%%%%%%%%%%%%%%%%%%%%%%%%%%%%%%%%%%%%%%

\begin{acknowledgments}
This work is based on observations made with the NASA/ESA/CSA James Webb Space Telescope.  The data were obtained from the Mikulski Archive for Space Telescopes at the Space Telescope Science Institute, which is operated by the Association of Universities for Research in Astronomy, Inc., under NASA contract NAS 5-03127 for JWST. These observations are associated with program \#2104. 
Based on observations made with the NASA/ESA Hubble Space Telescope, and obtained from the Hubble Legacy Archive, which is a collaboration between the Space Telescope Science Institute (STScI/NASA), the Space Telescope European Coordinating Facility (ST-ECF/ESAC/ESA) and the Canadian Astronomy Data Centre (CADC/NRC/CSA).
This research is based on observations made with the NASA/ESA Hubble Space Telescope obtained from the Space Telescope Science Institute, which is operated by the Association of Universities for Research in Astronomy, Inc., under NASA contract NAS 5–26555. 
This paper makes use of the following ALMA data: ADS/JAO.ALMA\#2017.1.00212.S. ALMA is a partnership of ESO (representing its member states), NSF (USA) and NINS (Japan), together with NRC (Canada), MOST and ASIAA (Taiwan), and KASI (Republic of Korea), in cooperation with the Republic of Chile. The Joint ALMA Observatory is operated by ESO, AUI/NRAO and NAOJ.  The National Radio Astronomy Observatory is a facility of the National Science Foundation operated under cooperative agreement by Associated Universities, Inc.  

\tmod{We thank the anonymous referee for the helpful comments that have improved this letter. }
DH is supported by Center for Informatics and Computation in Astronomy (CICA) grant and grant number 110J0353I9 from the Ministry of Education of Taiwan. DH also acknowledges support from the National Science and Technology Council of Taiwan through grant number 111B3005191. HC's research group is supported by an OPUS research grant (2021/41/B/ST9/03958) from the Narodowe Centrum Nauki. PB acknowledges the support of the Swedish Research Council (VR) through contract 2017-04924. The research of LEK is supported by a research grant (19127) from VILLUM FONDEN. JPR and ZYL are supported in part by NSF grant AST-1910106 and NASA grant 80NSSC20K0533. JPR would like to further acknowledge the support of the Virginia Initiative on Cosmic Origins (VICO). The authors also acknowledge financial support by NASA through a grant from the STScI.
\end{acknowledgments}

\vspace{5mm}
\facilities{JWST(NIRSPEC), ALMA, HST(NICMOS)}


\software{astropy \citep{2013A&A...558A..33A,2018AJ....156..123A}, numpy \citep{numpy},  matplotlib \citep{matplotlib}}



\appendix

\section{Centroid Alignment}\label{app:align}

\begin{figure*}
    \centering
    \includegraphics[width=0.95\textwidth]{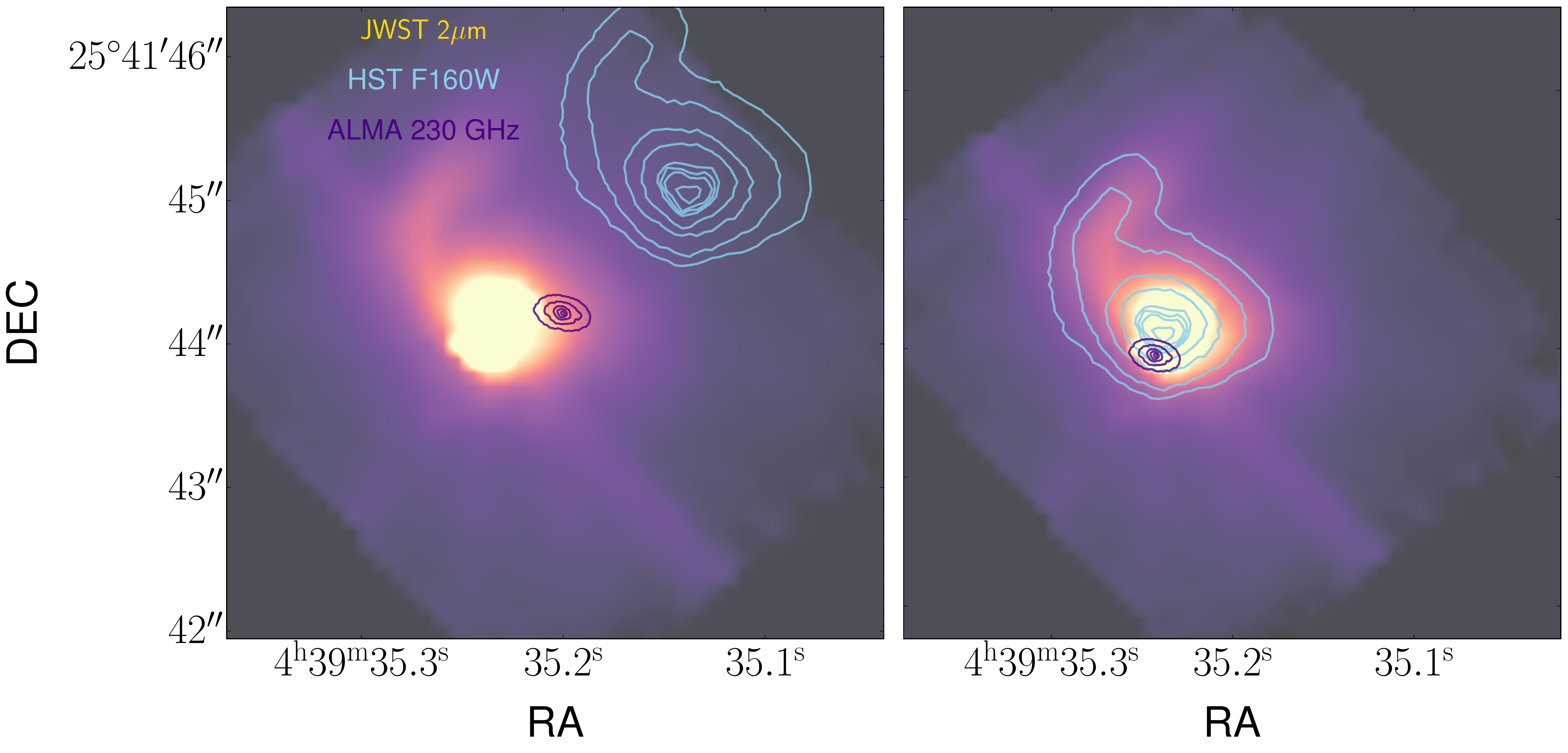} 
    \caption{{\it Left:} Original positions of the images. {\it Right:} Aligned images after correcting for proper motion.}
    \label{fig:align}
\end{figure*}

We have used the {\it HST} and ALMA images that were taken a few years apart to characterize the physical structure around TMC1A. The dithered HST images were processed in April 2005 using \emph{MULTIDRIZZLE} (v 3.1.0 Koekemoer). These data were directly downloaded from the MAST archive. The observations were taken in September 1997. The long-baseline ALMA data are published in \citet{bjerkeli2016}, and were taken in October 2017. Between when the HST, ALMA, and {\it JWST} were taken, TMC1A has moved considerably.

In order to align the images, we have used the arc structure that is seen in the {\it HST} image and the {\it JWST} dust continuum image.  Due to the positional uncertainty of HST NICMOS, the images were aligned with a shift that corresponds to a proper motion of ($\mu_{\alpha}$, $\mu_{\delta}$) = (0\farcs{036}/year, -0\farcs{026}/year) to align the {\it HST} and {\it JWST} images (see Fig.~\ref{fig:align}). The ALMA images are shifted by $\Delta \alpha =$0\farcs{6} and $\Delta \delta = $ -0\farcs{26} to align with the 2$\mu$m image.  We have used the long baseline ALMA data from 2016, 2017, and 2022 to confirm that the misalignment between {\it JWST} and {\it HST} is due to the positional offsets. We note that the ALMA data are not perfectly centered on the 2$\mu$m dust emission but it is located a few pixels to the southeast.

\section{VLT CRIRES vs {\it JWST}}\label{app:rovib}

\begin{figure}
    \centering
    \includegraphics[width=0.5\textwidth]{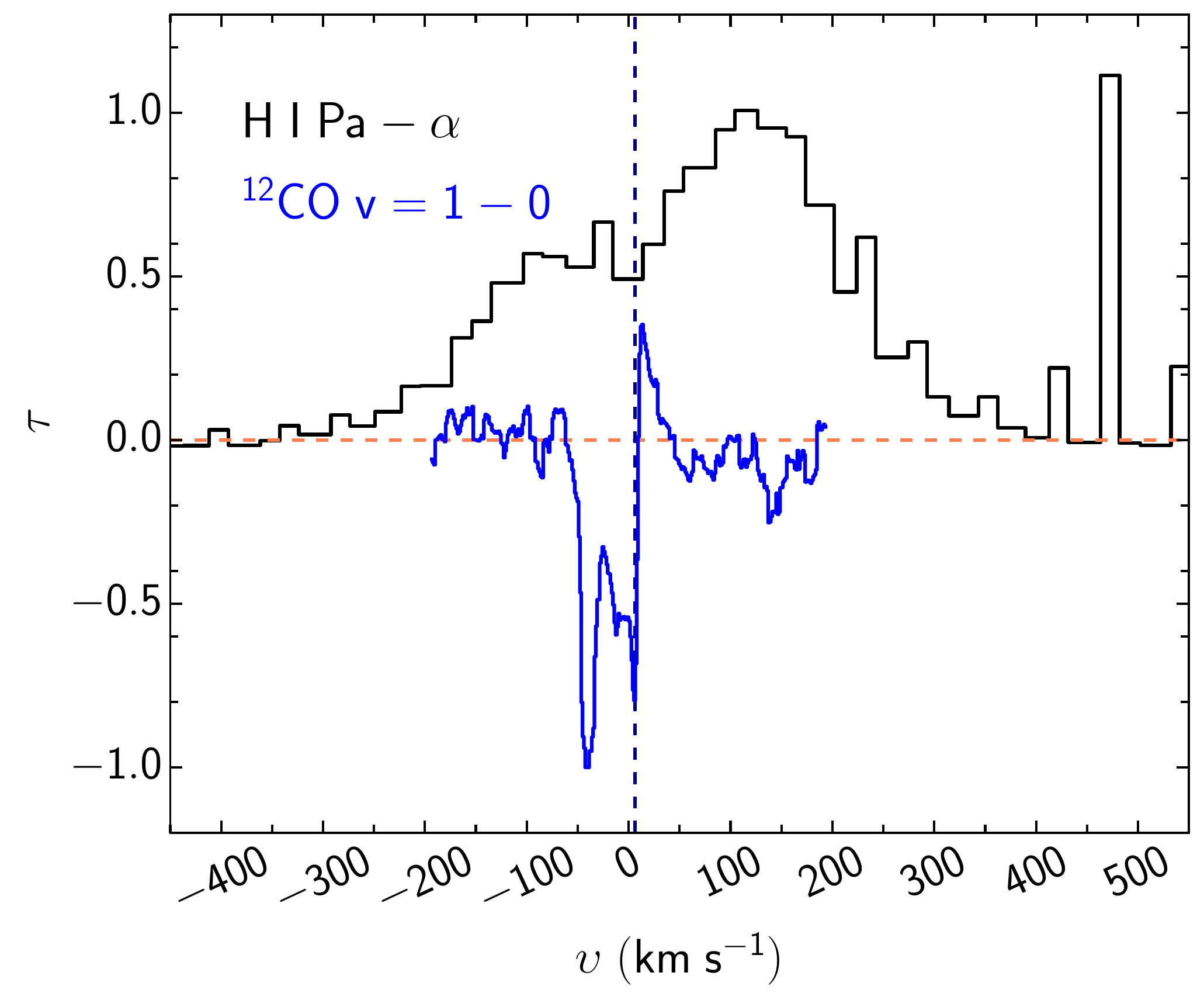} 
    \caption{Comparison between \ce{^{12}CO} (4.65 $\mu$m) and \ion{H}{1} (1.87 $\mu$m) normalized spectra. }
    \label{fig:12CO}
\end{figure}

Previous ground-based observations of ro-vibrational transitions of \ce{^{12}CO} toward TMC1A show a deep blue-shifted absorption and red-shifted emission \citep{herczeg11}. Figure~\ref{fig:12CO} shows the comparison between the {\it JWST} \ion{H}{1} Pa-$\beta$ and the averaged $v=1-0$ line of \ce{^{12}CO}. Given the overlap in velocity space, the material that is responsible for the strong absorption in \ce{^{12}CO} may also be responsible for the deficit of emission in the blue-shifted part of the \ion{H}{1} lines.

\bibliography{JWST1}{}
\bibliographystyle{aasjournal}



\end{document}